\documentclass[12pt,preprint]{aastex}
\input psfig.sty
\def\pn{{\par\noindent}}
\font\small = cmr7
\def\thn{{\thinspace}}
\def\sc{\scriptstyle}
\def\Msun{\hbox{$\thn M_{\odot}$}}
\def\Rsun{\hbox{$\thn R_{\odot}$}}

\def\={\thn\thn=\thn\thn}

\def\tgs{{\thn \rlap{\raise 0.5ex\hbox{$\sc  {>}$}}{\lower 0.3ex\hbox{$\sc  {\sim}$}} \thn }}
\def\tls{{\thn \rlap{\raise 0.5ex\hbox{$\sc  {<}$}}{\lower 0.3ex\hbox{$\sc  {\sim}$}} \thn }}
\def\tll{{\raise 0.3ex\hbox{$\sc  {\thn \ll \thn }$}}}
\def\tgg{{\raise 0.3ex\hbox{$\sc  {\thn \gg \thn }$}}}
\def\tle{{\raise 0.3ex\hbox{$\sc  {\thn \le \thn }$}}}
\def\tge{{\raise 0.3ex\hbox{$\sc  {\thn \ge \thn }$}}}
\def\tl{{\raise 0.3ex\hbox{$\sc  {\thn < \thn }$}}}
\def\tg{{\raise 0.3ex\hbox{$\sc  {\thn > \thn }$}}}
\def\ts{{\raise 0.3ex\hbox{$\sc  {\thn \sim \thn }$}}}

\def\do{{\tt "}}
\def\tp{{\raise 0.3ex\hbox{\small +}}}

\def\eq{{\thn\equiv\thn}}

\def\sep{{\par \noindent \hangindent=15pt \hangafter=1}}
\def\deg{{^\circ}}

\newcommand{\ergs}{erg~s$^{-1}$}

\newcommand{\msun}{$M_\sun$}
\def\alf{Alfv{\'e}n}
\def\pin{P_{\rm in}}
\def\ein{e_{\rm in}}
\def\ain{a_{\rm in}}
\def\pout{P_{\rm out}}
\def\prot{P_{\rm rot}}
\def\pkc{P_{\rm KC}}
\def\ttf{t_{\rm TF}}
\def\tv{t_{\rm visc}}

\shorttitle{X-Ray Sources in Multiple Stars}
\shortauthors{Makarov \& Eggleton}
\begin{document}

\title{The Origin of Bright X-Ray Sources in Multiple Stars} 
\author{V. V. Makarov \altaffilmark{1} \& P. P. Eggleton \altaffilmark{2}}
\affil{$^1$NASA Exoplanet Science Institute, California Institute of Technology, \\ Pasadena, CA 91125}
\affil{$^2$ Lawrence Livermore National Laboratory, Livermore, CA 94550}
\email{valeri.makarov@jpl.nasa.gov,eggleton1@llnl.gov}

\begin{abstract}
Luminous X-ray stars are very often found in visual double or multiple
stars. Binaries with periods of a few days 
possess the highest degree of coronal X-ray activity among regular, non-relativistic
stars because of their fast, tidally driven rotation. 
But the orbital periods in visual double stars are too large
for any direct interaction between the companions to take place. We suggest that most
of the strongest X-ray components in resolved binaries are yet-undiscovered short-period
binaries, and that a few are merged remnants of such binaries. The omnipresence 
of short-period active stars, e.g. of BY-Dra-type
binaries, in multiple systems is explained via the dynamical evolution of triple stars
with large mutual inclinations. The dynamical perturbation on the inner pair pumps up the
eccentricity in a cyclic manner, a phenomenon known as Kozai cycling. At times of close
periapsis, tidal friction reduces the angular momentum of
the binary, causing it to shrink. When the orbital period of the inner pair drops to a few days,
fast surface rotation of the companions is driven by tidal forces, boosting 
activity by a few orders of magnitude. If the period
drops still further, a merger may take place leaving a rapidly-rotating active dwarf
with only a distant companion.
\end{abstract}

\label{firstpage}
\section{Introduction}

\par An intriguing relation has been found between binarity and X-ray activity of normal stars.
For example, Makarov (2003) showed that among the 100 brightest X-ray sources within
$50$ pc of the Sun, a high proportion (79) are in binaries. RS CVn-type spectroscopic
binaries are by far the strongest X-ray sources among normal stars.
These binaries comprise at least one evolved (luminosity class IV or III) companion
and have orbital periods typically less than 10 days. 
Most of such objects
have X-ray luminosities in excess of $L_X=2.5\cdot 10^{30}$~\ergs. The origin of this
enhanced coronal X-ray radiation is commonly accepted to lie in
the high surface rotation rates supported by the orbital momentum and
tidal friction (Biermann \& Hall 1976). 

\vskip 0.2truein
\centerline{\psfig{figure=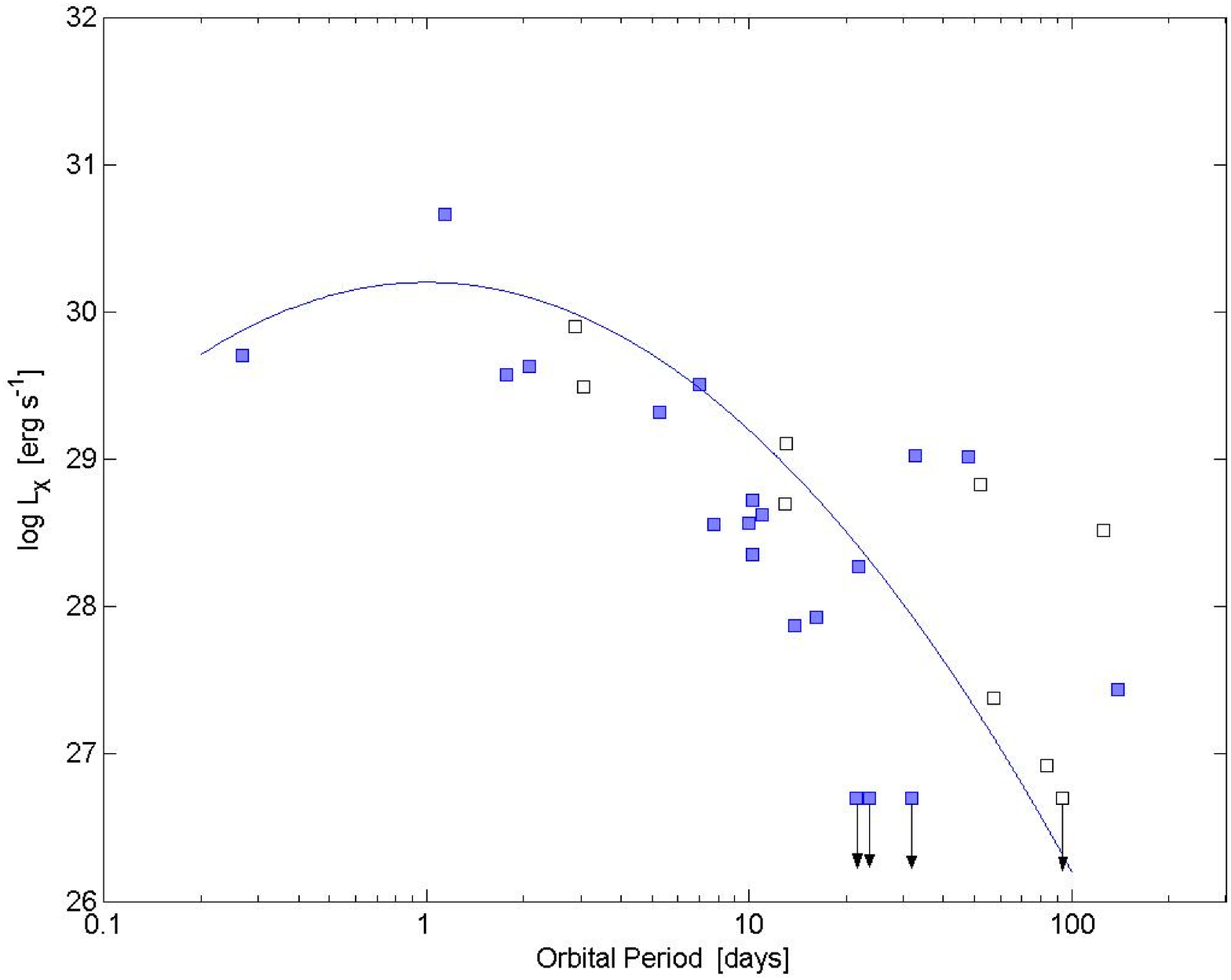,height=3.0in,bbllx=0pt,bblly=0pt,bburx=1250pt,bbury=850pt,clip=}}
\vskip -0.1truein
\pn {Fig 1 -- X-ray luminosities of all short-period solar-type 
spectroscopic binaries within 25 pc 
of the Sun. A polynomial fit is shown only for reference. 
Binaries with known tertiary or multiple 
companions are marked with filled squares.}

\par The graph in Fig. 1 shows X-ray luminosities of
all solar-type ($B-V \geq 0.4$ mag) spectroscopic binaries with distances less than $25\thn$pc
and periods less than 200 d listed in the SB9 catalog (Pourbaix et al. 2004), 
versus orbital periods in days. Objects not detected
by ROSAT are marked with downward arrows. This set includes binaries of RS CVn type
as well as main sequence binaries of BY Dra type. Although the scatter is
significant and the sample is small, the graph indicates that 
$L_X$ is correlated with the orbital period, and hence, with the rotation period
in late-type spectroscopic binaries. The highest
degrees of X-ray activity are found at periods between 1 and 10{\thn}d. A similar
relation between $\log L_X$ and rotation period was derived for late dwarfs 
by Pizzolato et al. (2003; their Fig. 3). As explained in Section~2,
a brightness-limited survey of X-ray sources, such as ROSAT, is dominated by the
most powerful emitters, those with $\log L_X$ well in excess of 29. In the Hyades open
cluster, such sources mostly reside in spectroscopic binaries with orbital periods
between 1 and {10\thn}d (Stern et al 1995).  RS CVn-type binaries, which include at least
one evolved component, are significantly brighter in X-rays than dwarf binaries, and the peak of their
$L_X$ distribution seems to be closer to 10 days (Makarov 2003). Singh et al. (1996) found
that about 45\% of field RS CVn-type objects they studied, had $\log L_X > 31$, but
only 15\% of Algol-type (semi-detached) binaries were as bright. One of the possible explanations 
for this fact is the enhancement of X-rays by the mechanism of reconnection of magnetic
lines around the detached RS CVn components, which are usually both active.
Interestingly, their Fig. 4a suggests that $L_X$ goes down at periods less than 1 d,
for these types of binaries, but Fig. 4b indicates that the surface flux rises up to the
shortest periods. Contact binaries of W UMa type are the shortest-period systems,
whose X-ray luminosities are mostly between $\log L_X=29$ and 30 (McGale et al. 1996),
noticeably weaker than the brightest BY and RS CVn type emitters (Dempsey et al. 1993).
These stars would occupy the area to the left and below the parabola's maximum
in our Fig.~1. 

The tendency of strong X-ray radiation to appear in short period (and therefore,
rapidly rotating) binary systems is well-documented and reasonbly well understood.
In this paper, we are concerned with the other part of the puzzle, namely, why
these enhanced emitters are often found in wide multiple systems. Indeed,
twenty out of a total 29 systems (67\%) are parts of hierarchical multiple systems 
in Fig.~1, indicated with filled
squares. The difference in $L_X$ distribution between single and binary
dK dwarfs with orbital periods longer than 1 yr in the Hyades is very pronounced, 
reaching almost an order of magnitude (Pye et al. 1994). Half of the high-fidelity
Hyades members with $\log L_X /\log L_{\rm Bol} > -4.8$ reside in visual binaries
with separations precluding any tidal interactions between the components (Makarov 2000).
These results imply that enhanced X-ray activity in normal stars occurs in short-period
binaries or rapidly rotating stars, which, in their turn, mostly appear in multiple systems.

\section{X-ray sources in visual binaries}

There is a strong tendency for the brightest X-ray sources detected by 
the ROSAT all-sky survey to reside in resolved visual double stars 
(Makarov 2002). This would not be surprising if one could extend the 
previous argument about the short-period binaries on to this case. 
However, many of these double stars are quite wide: they are visual or 
astrometric binaries, and there seems no obvious reason why these 
should be preferred. Components in wide binaries can be expected to be 
rapidly rotating if they are young, because young stars are usually
rapidly rotating, but we would not expect their frequency in {\it wide}
binaries to be higher than average. Besides, some of
the prominent X-ray stars are known to be sufficiently old,
such as the hard cataclysmic variable CQ Dra ($\log L_X=30.23$, HR$=1$)
orbiting an M giant with a period of 1703{\thn}d (Reimers et al. 1988).

The answer comes from a closer examination of the most prodigious X-ray 
emitters among the nearby short-period binaries. Most of the 
binaries represented in Fig. 1 are parts of wider, hierarchical multiple 
systems, as indicated with filled squares in this graph. If we 
limit ourselves to considering only binaries with periods less than 
$30\thn$d, 17 out of 21 binaries (81\%) belong to multiple systems, 
i.e. have wider companions. Although the sample is small, the result is 
statistically strong. The active binary TZ CrB (HR 6063, HIP 79607)
is a typical example of a very active, short-period ($P=1.14\thn$d), 
circularized binary in the solar vicinity. It is composed of two nearly 
identical solar analogs (masses 1.10 and 1.09 \msun), but it also has a 
wider companion of similar mass ($P=889\thn$yr, mass 1.06 \msun) according
to Tokovinin et al. (2006). The inner pair is a bright X-ray source with a 
count rate of $9.49$ counts per second as measured by Rosat. The tertiary 
companion is separated by $5.9\arcsec$ on the sky, and the distance to the 
system is approximately $20\thn$pc. A similar triple star
at $d=200\thn$pc would appear as a visual double with a separation of 
$0.6\arcsec$, which could be easily resolved either in Hipparcos 
(resolution limit $0.1\arcsec$; ESA 1997) or TDSC (resolution limit 
$0.3\arcsec$; Fabricius et al. 2002), the two large catalogs used 
by Makarov (2002) to identify double stars. The brightness of this system 
would be $0.095$ counts per second, well above the sensitivity limit of 
the Rosat Bright Source Catalog (RASS-BSC), $0.05$ counts per second 
(Voges et al. 1999). Even at a distance of $400\thn$pc, such a system 
would probably be detected in the Rosat Faint Source Catalog (RASS-FSC; 
Voges et al. 2000), and still be resolved by Hipparcos as a visual double star. This
example leads us to suspect that many of the double stars overluminous in X-rays are
in fact yet-undiscovered triple systems with short-period inner binaries. 

Assuming a typical hardness ratio for solar-type coronal sources of 
HR1$=-0.2$, we estimate that 1-day binaries at $L_X\approx 10^{30}\thn$\ergs 
would be detected and listed in RASS-BSC out to $150\thn$pc. The detection 
radius for a 10-day binary at  $L_X\approx 10^{29}\thn$\ergs would be
approximately $50\thn$pc. Evolved binaries of RS CVn type are considerably more luminous in
X-rays than dwarfs, and can often be detected out to $300-400\thn$pc. Therefore, an X-ray
brightness-limited sample of nearby stars will comprise a large fraction of short-period
binaries. The remaining difficulty is to explain why these binaries prefer to be
accompanied by wider tertiary companions of similar mass.

\par Recently Eggleton \& Kisseleva-Eggleton (2006; EKE06) suggested that 
AB~Dor, one of the sample in Makarov (2003), had a history that involved 
(a) an initial multiplicity higher than two, and (b) a recent merger of 
two components of a {\it close} binary that now no longer exists except 
as a single merged remnant. In fact AB~Dor is the brightest component, 
visually as well as in X-rays, of a quadruple system, but for present 
purposes we disregard the {\it distant} companion Rst 137B (AB~Dor~B), which 
is itself a $0.07\arcsec$ binary (Janson et al. 2007) at $\sim 9\arcsec$ 
from the principal component AB~Dor~A (Innis et al. 1986); and so we consider 
only the astrometric sub-binary AB~Dor~AC, with period $11.74\thn$yr 
(Nielsen et al. 2005). AB~Dor~A is a very active and rapidly rotating K 
dwarf ($0.51\thn$d, Innis et al. 1988), and its X-ray brightness and high 
level of activity is no doubt due to this rapid rotation. AB~Dor~C is a very 
late (M7 -- M9) dwarf of $0.09\Msun$, near the very bottom of the 
hydrogen-burning Main Sequence (MS);
it is sufficiently near the MS that it cannot be extremely young. Nielsen et al.
(2005) estimated the age as $70\pm 30\thn$Myr.
It is not clear 
how AB~Dor~A, rotating near to its 
break-up rate, and subject to the mass ejection and the magnetic braking that
is consequent on that spin, has maintained such {\it extremely} rapid rotation while most 
other stars of comparable age have braked considerably.

\par EKE06 suggested a prior evolution of AB~Dor~AC of the following form.
The $11.74\thn$yr binary was once a triple system, but a close pair within
this triple was driven, rather recently, into a merger. The present AB~Dor~A
is the rapidly rotating remnant of that merger. If the merger took place
only perhaps $1-5\thn$Myr ago, we can understand why it is still rotating at
close to break-up. The previous, now merged, close binary might have had an
orbital period at birth of several days, or even weeks or months, and would have been 
driven into a merger by a combination of Kozai cycling and tidal friction 
(KCTF; Mazeh \& Shaham 1979, Kiseleva et al. 1998) that was dictated by 
AB~Dor~C, the companion in the $11.74\thn$yr binary. Magnetic braking 
with tidal friction (MBTF; Huang 1966, Mestel 1968, van't Veer 1976, Rucinski 
1983) will probably also have played a part in driving this sub-sub-binary 
to a merger. 

\par The ultrafast rotator Speedy Mic (BO Mic, HIP 102626) is another example of an
active isolated star in the solar neighborhood, whose origin remains a mystery.
This early K dwarf with a rotational period of 0.38$\thn$d (Cutispoto et al. 1997) 
and $\log L_X = 31.07$ is
not represented in Fig. 1, because there is no documented evidence, to our knowledge,
of spectroscopic binarity. It has been known, however, since the publication of the 
Hipparcos catalog, as an astrometric binary with changing proper motion 
(ESA 1997). The invisible companion may be responsible for the outstanding properties
of BO Mic, if it caused the original inner pair to merge recently, similar to the scenario
discussed above for AB Dor. The orbital period of the companion is of particular interest,
because BO Mic is suspected to be quite young, possibly a post-T Tauri star. If the orbital
period is long, our surmise is refuted, because the KCTF did not have enough time
to merge the inner pair. We have attempted unconstrained orbital solutions based on
the Hipparcos Intermediate Abscissae Data, using the method of Goldin \& Makarov (2006).
In this case again, we encountered the persistent problem of unconstrained astrometric
solutions on undersampled data, that the orbital eccentricity can not be determined to the 
same relative precision as, for example, the period. Two alternative solutions were 
therefore produced, one with
unconstrained eccentricity, and the other with eccentricity set to zero. The former
solution is: period $P=1146\thn$d, inclination $i=78 \deg$, eccentricity $e=0.63$,
apparent semimajor axis $a_0=24.0\thn$mas, and parallax $\Pi=25.0\thn$mas. The latter 
solution
is $P=1213\thn$d, $i=82 \deg$, $e=0.00$, $a_0=18.9\thn$mas, and $\Pi=23.1\thn$mas. 
Either of these solutions implies that the companion's
mass is not much smaller than the primary, because the apparent orbital motion is slightly
less than 1 AU, and the relative semimajor axis is slightly larger than 2 AU. The
orbital period is short enough to make the KCTF merging scenario plausible.

\section{Kozai Cycles, Tidal Friction and Magnetic Braking }
\par The essential points about Kozai cycling, tidal friction and magnetic braking, 
are:
\sep (a) If the inner orbit (with period $\pin$) of a triple is inclined at 
an angle $\eta$ to the outer orbit (period $\pout$), and if $\sin\eta\tge\sqrt{2/5}$
($39\deg\tl\eta\tl 141\deg$), then the inner orbit is forced to cycle 
in eccentricity and in inclination, but {\it not} in period or semimajor
axis, on a Kozai-cycling period $\pkc$; see point (c) below. The amplitude of
the eccentricity cycle depends {\it only} on (i) the minimum eccentricty, 
and (ii) the inclination $\eta$. It does {\it not} depend on either the 
mass of the distant third body or its distance (i.e. on $\pout$). The 
range as a function of inclination and minimum eccentricity has been 
tabulated by Eggleton (2006; his Table 4.9). For instance if 
$\eta=60\deg$ the eccentricity can range between zero and 0.764, or 
between 0.5 and 0.863. An inclination of $60\deg$ is the median to be 
expected if outer and inner orbits are oriented at random. 

\sep (b) However, purest Kozai cycling requires that gravity be exactly 
inverse-square-law. Modifications of gravity due to (i) General Relativity,
(ii) the quadrupole moment of an extended star that rotates, and (iii)
the quadrupole moments of each component due to the attraction of the other,
can all destroy Kozai cycles, if these modifications are large enough.
Very loosely, if $\pout{\rm(yrs)}\tgs\left[\pin{\rm(dys)}\right]^{1.4}$,
the effect of Kozai
cycling is suppressed. It is of course also suppressed at an opposite
extreme: if $\pout$(yrs)$/\pin$(yrs)$\tls 4$, the inner orbit is likely to 
be so large relative to the outer orbit that the system is destroyed 
dynamically within a short time. For example, with $\pin\ts 16\thn$d,
Kozai cycling can take place if $0.2\thn$yr$\tls\pout\tls 50\thn$yr.

\sep (c) The duration of a Kozai cycle is roughly
$\pkc\ts {\pout^2/\pin}\thn\cdot\thn (m_1+m_2+m_3)/m_3\thn.$
Thus Kozai cycling can take place on timescales as short as a few centuries,
but more typically on timescales of Kiloyears to Megayears, and even
Gigayears. With 
$\pin\ts 16\thn$d, $\pout\ts 12\thn$yr and $m_3\ts 0.1(m_1+m_2)$,
the Kozai cycle period is of order $3\times10^4\thn$yr.

\sep (d) To a good degree of approximation the outer orbit is
not affected at all by Kozai cycling.

\sep (e) Because the eccentricity $\ein$ can become large during a cycle while the
semimajor axis $\ain$ is constant, the periastron separation $p\eq \ain(1-\ein)$ can 
become small, and as a result tidal friction can become an important dissipative 
agent. The dissipation of energy reduces $\ain$ and $\pin$, just as in normal
close but eccentric binaries. But in a purely binary situation the
decrease of $\ain$ is limited by the fact that the orbit becomes circular 
and so dissipation stops: the final $\ain$ is roughly the initial semi-latus-rectum.
In the triple situation there is no such limitation, and so we might
expect $\pin$ to decrease indefinitely. However, this is limited by~(b)~above,
so that the final product has a circular inner binary that no longer Kozai cycles, 
but is of sufficiently short period that either (i) the two components distort each other
significantly, or (ii) their spins, in corotation with their orbit, cause
enough of a quadrupole moment to suppress the Kozai cycling. In principle,
the apsidal motion of GR might also do this, but in practice the other two effects
are more important at short $\pin$.

\sep (f) The timescale $\ttf$ of tidal friction can be estimated
(Zahn 1977) as $\ttf\ts\tv \thn(p/R)^8$, where $R$ is the radius of the larger
star; $\tv$ is a viscous timescale inherent in the structure of the star,
and is $\ts 1\thn$yr. The fact that most binaries with periods $\tls 4\thn$d 
have small or 
zero eccentricity is seen as confirmation that the circularisation timescale 
is of the order of the nuclear timescale when the orbital radius is about
ten times the stellar radius. In some old clusters, and the Galactic halo, 
orbits with periods
as long as $10\thn$d or even longer may be circularised (Latham et al. 1992).

\sep (g) If the components of the close pair are of spectral type F -- M,
and if its period becomes short ($\tls 10\thn$d) as a result of KCTF, then
orbital shrinkage will continue under the influence of magnetic braking, also
in association with tidal friction -- MBTF. F/G/K/M stars that rotate in 
$\prot\tls 10\thn$d are usually very X-ray active (Fig. 1) and this is 
normally attributed to dynamo activity driven by the combination of relatively
rapid rotation and a turbulent surface convective zone. The dynamo activity
means both that a stellar wind is driven off by the dissipation of magnetic
field as it emerges just outside the stellar surface, and that this wind
carries off further magnetic field. The magnetic field causes the wind to 
corotate with the star out to an \alf\ radius of several stellar radii
(about $10\Rsun$ in case of the Sun itself), and thus carries off a great deal
of angular momentum which, beyond the \alf\  radius, is lost to the star.
In a single star this would simply cause the star to spin more slowly,
but in a binary whose components are close enough to be locked in corotation
with the orbit by tidal friction this would cause the {\it orbit} to lose
angular momentum and thus to spin {\it faster}. It is therefore an unstable
process, although even as it runs away, with the period decreasing until
the larger star is in contact with its Roche lobe at $\ts 0.3\thn$d (for
the Sun), the timescale is unlikely to be much smaller than about $10\thn$Myr.

\sep (h) KCTF can spin the close sub-binary up to $\pin \tls 3\thn$d, and then
MBTF can spin it up further, by way of a very close but detached binary to 
Roche-lobe overflow (RLOF). This can lead first to a semi-detached `reverse 
Algol', where the loser is still the more massive component, and then
probably either (i) to a normal (but very short-period) Algol
or (ii) to a contact binary. Contact binaries are likely
to evolve by net mass transfer from the less to the more massive component,
until the less massive component is merged into its companion (Webbink 1976). 
Such a process is likely to be quite slow, however, since contact binaries
are by no means uncommon. But a further possibility (iii) is a {\it rapid} 
merger, on something like a timescale of years, and at its peak, days.
Such a merger is to be expected if the mass ratio at the onset of RLOF is
fairly extreme, like 2:1, 3:1 or even more; only if the mass ratio is rather
mild, say $\tls 1.5:1$, is there likely to be a settling-down to a relatively
long-lived semi-detached or contact state as in (i) or (ii). These evolutionary 
possibilities were discussed at some length by Yakut \& Eggleton (2005).

\par We illustrate the effect of KCTF and MBTF with an example that might be
relevant to the AB~Dor AC system. Consider a triple whose initial
parameters, in an obvious notation, are
\pn
\centerline{((K5V+M7V; $0.65+0.15\Msun$; $\pin=16\thn$d, $\ein=0$)\hskip 2truein} 
\pn
\centerline{\hskip 2truein + M8V; $0.8+0.09\Msun$; $\pout=11.74\thn$yr, $e=0.61$; $\eta = 84.3\deg$)\ .}
\pn Since $\cos\eta=0.1$, the probability is 10\% that the inclination will be this
large or larger, if the inclination is selected randomly because of some random
collision of two primordial binaries. The initial rotational period of
both components in the inner pair was set arbitrarily at $5\thn$d. We computed the 
subsequent evolution according 
to the prescriptions of Eggleton (2006) for KCTF and MBTF.


\vskip 0.2truein
{\centerline{\psfig{figure=pgplote1.ps,height=3.0in,bbllx=0pt,bblly=375pt,bburx=400pt,bbury=750pt,clip=}\hskip 3.5truein}
\vskip -3.0truein
{\centerline{\hskip 1.3truein\psfig{figure=pgplote1.ps,height=3.0in,bbllx=400pt,bblly=375pt,bburx=600pt,bbury=750pt,clip=}}
\vskip -3.0truein
{\centerline{\hskip 4.3truein\psfig{figure=pgplote2.ps,height=3.0in,bbllx=400pt,bblly=375pt,bburx=600pt,bbury=750pt,clip=}}
{Fig 2 -- Behaviour of eccentricty, inclination and period for the inner
binary of a triple proto-AB-Dor system. Left-hand panel: eccentricity as a
function of time. The periodic (Kozai cycling) behaviour at early times is
severely undersampled. Second panel: mutual inclination as a function of
time. Third panel: $\pin$, dark blue; $\prot$, pale blue; $\pout$, green.
See text for details. Right-hand panel: as third panel, but on a longer
timescale, so that the effect of MBTF leading to a merger at very short
period is now visible.}}
\vskip 0.2truein

\par The first two panels of Fig. 2 show the evolution of the eccentricity and 
the inclination over an interval of about $2.5\thn$Myr. In the first $0.1\thn$Myr the 
eccentricity (first panel) cycles over a considerable range, but the cyclic 
character of its variation is not easy to see on the scale plotted. The evolution
is severely undersampled so that the variation of $\ein$ looks random. Tidal
friction reduces the range of variation of $\ein$, until after $\ts 1.5\thn$Myr
the eccentricity fluctuates only slightly about a value of $\ts0.75$. The inclination
(second panel) also cycled originally, between about $77.3\deg$ and $84.3\deg$,
but settled down to a much narrower range about $78.2\deg$ by about $1.5\thn$Myr.
Subsequently the eccentricity decreased while only fluctuating slightly, and
halved itself roughly every $3\thn$Myr.  The third panel of Fig. 2 shows the inner
orbital period and the rotation period of the K5 dwarf, for the first $2.5\thn$Myr; 
and the fourth shows the same but for $\ts 60\thn$Myr. KCTF is largely over after 
$\ts 10\thn$Myr, with $\pin$ decreased to the value where the perturbative effects 
of (b) kill the cycling. But the period continues to decrease slowly because of 
MBTF, and then rather rapidly, presumably towards a merger, once $P\tls 0.5\thn$d 
at $60\thn$Myr.

\par  While the inner binary is still unmerged, it will be quite active (Fig.~1)
because its period is $\tls10\thn$d. But after the merger its
rotational period will be $\ts 0.25\thn$d, and now {\it increasing} as a result
of magnetic braking without tidal friction. The timescale for doubling the
period will be about $20\thn$Myr -- the code used did not simulate either the
merger or the later evolution.
\section{Discussion}
We interpret the abundance of wide double stars among the brightest X-ray sources
as a direct consequence of the Kozai interaction in hierarchical triple system,
prerequisite to the generation of inner binaries with orbital periods less than 3 days.
Can we find observational evidence in support of this hypothesis? A large-scale
spectroscopic survey of randomly selected X-ray double stars would give a direct
proof. Currently, only a few selected stars have been investigated in depth, for
example, TZ CrB and AB Dor discussed in Section 2. But the main predictions of
this model are supported by theory and a variety of astronomical observations:
\begin{itemize}
\item
The strongest X-ray sources among normal stars are binaries with a period of a few days.
\item
Short-period binaries are mostly found in hierarchical triple or multiple systems.

\item
The time-scales of tidal evolution and Kozai cycling in triple systems of high mutual
inclination can be much shorter than the stellar evolution time-scale.

\item 
Torres et al.~(2003) examined 10 stars suspected to be members of the pre-main-sequence 
TW Hya association (TWA).
The candidate members were selected by high X-ray luminosities and kinematic or positional
similarity to known TWA members. Despite this pre-selection, upon spectroscopic and radial 
velocity measurement none of the examined objects turned out to be young, but no less
than six of them  were confirmed as binaries. Even more strikingly, all binaries whose
orbits could be determined had periods of less than 3 days. This result perhaps indicates
that the overwhelming majority of bright X-ray sources in the field are short-period
binaries, though most of them are yet to be discovered. 

\item 
Tokovinin et al.~(2006) looked at 165 known spectroscopic binaries with 
$P\tl 30\thn$d. Some of these already had known third (or more) companions, 
but they also found 13 new ones by NACO adaptive optics. Making allowance by 
a maximum-likelihood algorithm for observational incompleteness, they found 
that the shortest-period spectroscopic binaries, those with $P\tl 2.9\thn$d, 
were almost always (96\% likelihood) in triples, while the longest-period 
spectroscopic binaries, those with $12 - 30\thn$d, were substantially less 
likely (34\%).

\item 
Pribulla \& Rucinski (2006) found that for all (88) known contact binaries in
the Northern sky down to $V=10$, 52 (59\%) are in triples; Rucinski et al (2007) increased
that to 54 (61\%) with two new adaptive-optics discoveries. This is a much higher
proportion of triples than in a broad sample of binaries not confined to periods as short
as in contact binaries ($\tl 1\thn$d). Given the many selection effects that make the
discovery of faint companions difficult, this result (61\%) is not inconsistent
with 100\% in practice.
\item 

Eggleton \& Tokovinin (2008) considered the multiplicities of the 4559 stellar
systems brighter than Hipparcos magnitude 6.0. They found that 1841 were at least binary, 
and of these 404 were at least triple, i.e. about 22\% of binaries have a further 
companion. However, among the 1841 that were at least binary, the 89 that had a period 
$\tl 3\thn$d contained 55 (62\%) that were at least triple. 

\item 
Nuclear evolution may play a role in the evolution of some X-ray sources, 
specifically
the RS Cvn type. Although some of these have quite short periods, such as RS CVn itself
($4.8\thn$d), others have longer periods, e.g. RU Cnc ($10.17\thn$d) and RZ Eri 
($39.28\thn$d). Both contain giants, and it is presumably the fact that the giant is
fairly close to its Roche lobe, and forced by tidal friction to spin much more rapidly
than it would have done if it had evolved as a single star, that makes it X-ray active.
Probably KCTF and MBTF are irrelevant in these cases.
\end{itemize}

\pn The above items seem consistent only with the concept that most, and arguably all,
binaries with $P\tls 3\thn$d have a third body, and therefore that the third body must be 
essential for producing most or all of the short periods. This can only mean the KCTF
mechanism; although MBTF can shorten periods from 5 or maybe 10 days to one day or less,
there is no reason why this should involve a third body. Only one other mechanism might
be worthy of consideration: within a cluster, gravitational scattering of hard binaries by
other members of the cluster can harden the binaries, though generally by a rather small
amount in each scattering. But once again there is no particular reason why the hardest 
binaries should end up as members of triples.
\par In summary, KCTF along with other physical processes can account for the observed 
properties of many X-ray sources, and perhaps all, or at least the great majority, with 
short
periods. It can also account, through the merger mechanism described above specifically 
in the context of AB~Dor, for X-ray sources in {\it wide} binaries. It is quite possible
that luminous X-ray sources in wide binaries are themselves in close binaries that have
not yet been measured, and it is also possible that some are the merged remnants
of close binaries. They need not be rotating as rapidly as AB~Dor, which will probably
be conspicuous for several hundred Myr yet until it has slowed down to perhaps
$\prot\ts 12\thn$d; but they could in principle be in wide binaries and have ages
of several Gyr, having merged only after a substantially slower KCTF involving
different initial conditions.

\acknowledgments
The research described in this paper was carried out partly at the Jet Propulsion 
Laboratory, California Institute of Technology, under a contract with the National 
Aeronautics and Space Administration. 
This work was also performed partly under the auspices of the U.S. Department of 
Energy by Lawrence Livermore National Laboratory under Contract 
DE-AC52-07NA27344. This research has made use of the SIMBAD database,
operated at CDS, Strasbourg, France; and data products from the 2MASS, which is a 
joint project of the University of Massachusetts and the Infrared Processing and 
Analysis Center, California Institute of Technology, funded by NASA and the NSF.

\end{document}